\documentclass[pre,aps,amssymb,amsmath,superscriptaddress,twocolumn]{revtex4}
\usepackage{amsmath}
\usepackage{amssymb}
\usepackage{epsfig}
\usepackage{amscd}
\usepackage{graphicx,psfrag,xspace}
\usepackage{float}
\usepackage[normalem]{ulem}

\begin{document}

\title{Collision densities and mean residence times for $d$-dimensional exponential flights}
\author{A. Zoia}
\email{andrea.zoia@cea.fr}
\affiliation{Commissariat \`a l'Energie Atomique et aux Energies Alternatives, Direction de l'Energie Nucl\'eaire, D\'epartement de Mod\'elisation des Syst\`emes et Structures, Service d'Etudes des R\'eacteurs et de Math\'ematiques Appliqu\'ees, CEA/Saclay, 91191 Gif-sur-Yvette, France}
\author{E. Dumonteil}
\affiliation{Commissariat \`a l'Energie Atomique et aux Energies Alternatives, Direction de l'Energie Nucl\'eaire, D\'epartement de Mod\'elisation des Syst\`emes et Structures, Service d'Etudes des R\'eacteurs et de Math\'ematiques Appliqu\'ees, CEA/Saclay, 91191 Gif-sur-Yvette, France}
\author{A. Mazzolo}
\affiliation{Commissariat \`a l'Energie Atomique et aux Energies Alternatives, Direction de l'Energie Nucl\'eaire, D\'epartement de Mod\'elisation des Syst\`emes et Structures, Service d'Etudes des R\'eacteurs et de Math\'ematiques Appliqu\'ees, CEA/Saclay, 91191 Gif-sur-Yvette, France}

\begin{abstract}
In this paper we analyze some aspects of {\em exponential flights}, a stochastic process that governs the evolution of many random transport phenomena, such as neutron propagation, chemical/biological species migration, or electron motion. We introduce a general framework for $d$-dimensional setups, and emphasize that exponential flights represent a deceivingly simple system, where in most cases closed-form formulas can hardly be obtained. We derive a number of novel exact (where possible) or asymptotic results, among which the stationary probability density for $2d$ systems, a long-standing issue in Physics, and the mean residence time in a given volume. Bounded or unbounded, as well as scattering or absorbing domains are examined, and Monte Carlo simulations are performed so as to support our findings.
\end{abstract}
\maketitle

\section{Introduction}
\label{introduction}

Random walks are widely used in Physics so as to model the features of transport processes where the migrating (possibly massless) particle undergoes a series of random displacements as the effect of repeated collisions with the surrounding environment~\cite{hughes, weiss, feller, klafter}. While much attention has been given to random walks on regular Euclidean lattices, and to the corresponding scaling limits, less has been comparatively devoted to the case where the direction of propagation can change continuously at each collision: for an historical review, see, e.g.,~\cite{dutka}. Such processes, which are intimately connected to the Boltzmann equation, have been named {\em random flights}, and play a prominent role in the description of, among others, neutron or photon propagation through matter~\cite{bell, wigner, cercignani}, chemical and biological species migration~\cite{bacteria}, or electron motion in semiconductors~\cite{jacoboni_book}.

Within the simplest formulation of this model, which was originally proposed by Pearson (1905)~\cite{pearson} and later extended by Kluyver (1906)~\cite{kluyver} and Rayleigh (1919)~\cite{rayleigh}, it is assumed that particles perform random displacements (`flights') along straight lines, and that at the end of each flight (a `collision' with the surrounding medium) the direction of propagation changes at random.

When the number of transported particles is much smaller than the number of the particles of the interacting medium, so that inter-particles collisions can be safely neglected, it is reasonable to assume that the probability of interacting with the medium is Poissonian. For the case of neutrons in a nuclear reactor, e.g., the ratio between the number of transported particles and the number of interacting nuclei in a typical fuel$/$moderator configuration is of the order of $10^{-11}$, even for high flux reactors~\cite{cercignani}. It follows that flight lengths between subsequent collisions are exponentially distributed (hence we will call this process {\em exponential flights} in the following~\footnote{Remark that exponential flights are preferentially called {\em free flights} in the semiconductors community~\cite{jacoboni_book}.}). We assume that collisions can be either of scattering or absorption type. At each scattering collision, the flight direction changes at random, whereas at absorption events the particle disappears and the flight terminates. Each flight can be seen as a random walk in the phase space of position ${\mathbf r}$ and direction ${\mathbf \omega}$ in a $d$-dimensional setup.

The particle density $\Psi({\mathbf r},{\mathbf \omega},t)$ represents the probability density of finding a transported particle at position ${\mathbf r}$ having direction ${\mathbf \omega}$ at a time $t$, up to an appropriate normalization factor. In many applications, the actual physical observable is the average of the density $\Psi({\mathbf r},{\mathbf \omega},t)$ over the directions ${\mathbf \omega}$, namely
\begin{equation}
\Psi({\mathbf r},t) = \frac{1}{\Omega_d} \int \Psi({\mathbf r},{\mathbf \omega},t) d{\mathbf \omega},
\end{equation}
where $\Omega_d = \int d{\mathbf \omega} = 2 \pi^{d/2}/\Gamma\left(d/2 \right)$ is a normalization factor corresponding to the surface of the unit $d$-dimensional sphere and $\Gamma()$ is the Gamma function~\cite{abramowitz}.

Along with the development of Monte Carlo methods, numerical solutions $\Psi({\mathbf r},t)$ to complex three-dimensional linear/nonlinear transport problems coming from applied sciences are becoming accessible to an high degree of accuracy: criticality calculations in reactor cores~\cite{lux}, scattering and absorption in heated plasmas~\cite{tokamaks}, propagation through anisotropic scattering centers in atmosphere or fluids~\cite{anisotropy}, and charge transport in semiconductors under external fields~\cite{kurosawa, rev_elec_1, rev_elec_2}, only to name a few. Nonetheless, even for the simplest systems, many theoretical questions remain without an answer, so that the study of exponential flights has attracted a renovated interest in recent years; see, e.g.,~\cite{paasschens, orsingher, kolesnik, lecaer}. In particular, it has been emphasized that the dimension $d$ deeply affects the nature of $\Psi({\mathbf r},t)$, and prevents in most cases from obtaining explicit results. The aim of our work is to investigate exponential flights in a generic $d$-dimensional setup, under simplifying hypotheses. Here, we will mostly focus on establishing insightful relationships between space, time and the statistics of particle collisions within a given volume. A number of new results will be derived, concerning unbounded, bounded, scattering as well as absorbing domains.

This paper is structured as follows. In Sec.~\ref{setup} we recall the mathematical formalism, introduce the physical variables and derive their inter-dependence for any $d$. In Sec.~\ref{space_moments} we detail the structure of the spatial moments of the particle ensemble. Sec.~\ref{coll_density_sec} is devoted to the analysis of the collisions statistics in a given domain. Then, in Sec.~\ref{dimensional_analysis} we examine the distinct cases $d=1,2, 3$ and $4$. Both the spatial and temporal evolution of the particle ensemble are considered, and results for bounded domains are obtained by resorting to the method of images. We provide a comparison between analytical (or asymptotic) findings and Monte Carlo simulations. Conclusions will be finally drawn in Sec.~\ref{conclusions}.

\begin{figure}[t]
   \centerline{ \epsfclipon \epsfxsize=9.0cm
\epsfbox{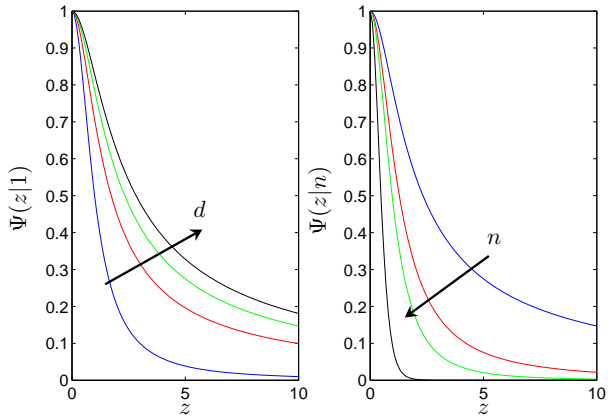} }
   \caption{The free propagator $\Psi(z|n)$ in the transformed space. Left: $\Psi(z|1)$ for increasing values of the dimension, $d=1$ (blue), $2$ (red), $3$ (green), and $4$ (black). Right: $\Psi(z|n)$ for $d=3$ and increasing number of collisions, $n=1$ (blue), $2$ (red), $3$ (green), and $10$ (black).}
   \label{fig1}
\end{figure}

\section{General setup}
\label{setup}

Within the natural framework of statistical mechanics, the evolution of the particle density $\Psi({\mathbf r},{\mathbf \omega},t)$ for exponential flights is governed by the linear Boltzmann equation~\cite{cercignani}. Linearity stems from neglecting inter-particle collisions. In the hypothesis that an average particle energy can be defined (the so called one-speed transport), and that the physical properties of the medium do not depend on position nor time, the Boltzmann equation for the density $\Psi({\mathbf r},{\mathbf \omega},t)$ reads~\cite{cercignani, paasschens}
\begin{eqnarray}
\frac{1}{v}\frac{\partial}{\partial t} \Psi({\mathbf r},{\mathbf \omega},t) + {\mathbf \omega} \cdot \nabla_{\mathbf r} \Psi({\mathbf r},{\mathbf \omega},t) =\nonumber \\
= -\sigma_t\Psi({\mathbf r},{\mathbf \omega},t)+\sigma\int k({\mathbf \omega'},{\mathbf \omega}) \Psi({\mathbf r},{\mathbf \omega'},t)d{\mathbf \omega'} + \frac{{\cal S}}{v},
\label{boltzmann}
\end{eqnarray}
where $\sigma_t$ is total cross section of the traversed medium (carrying the units of an inverse length), $\sigma$ is the scattering cross section, $v$ is the particle speed, and ${\cal S}$ is the source. The total cross section $\sigma_t$ is such that $1/\sigma_t$ represents the average flight length between consecutive collisions (the so-called mean free path), and is related to the scattering cross section $\sigma$ and to the absorption cross section $\sigma_a$ by $\sigma_t=\sigma+\sigma_a$. The quantity $k({\mathbf \omega'},{\mathbf \omega})$ is the scattering kernel, i.e., the probability density that at each scattering collision the random direction changes from ${\mathbf \omega'}$ to ${\mathbf \omega}$.

\begin{figure}[t]
   \centerline{ \epsfclipon \epsfxsize=9.0cm
\epsfbox{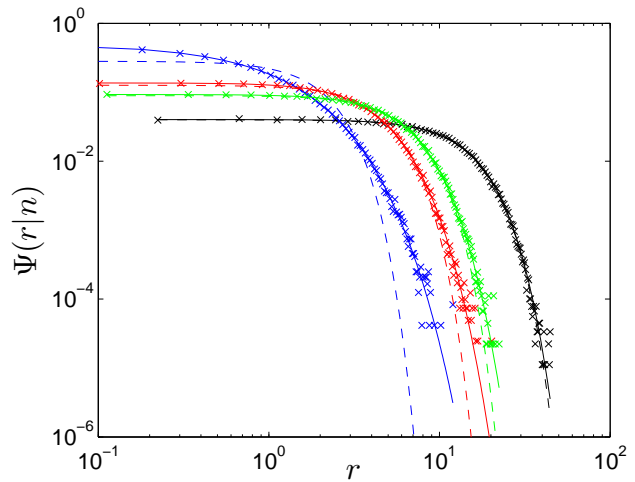} }
   \caption{The free propagator $\Psi(r|n)$ for $d=1$, with $n=1$ (blue), $5$ (red), $10$ (green), and $50$ (black). Monte Carlo simulation results are displayed as symbols. The solid line is the theoretical result, Eq.~\eqref{1d_propagator}. The dashed line is the diffusion limit, Eq.~\eqref{diff_eq_n}.}
   \label{figpn_d1}
\end{figure}

Denoting by $\Psi({\mathbf r},{\mathbf \omega},t)$ the solution of the Boltzmann equation~\eqref{boltzmann} for a medium without absorptions ($\sigma_a = 0$), the complete solution with absorption $\Psi_a({\mathbf r},{\mathbf \omega},t)$ can be easily obtained by letting
\begin{equation}
\Psi_a({\mathbf r},{\mathbf \omega},t) = \Psi({\mathbf r},{\mathbf \omega},t) e^{-v \sigma_a t},
\label{boltzmann_absorption}
\end{equation}
thanks to linearity~\cite{paasschens}. This allows primarily addressing a purely scattering medium ($\sigma_t = \sigma$), without loss of generality.

At long times, i.e., far from the source, the direction-averaged particle density $\Psi({\mathbf r},t)$ is known to converge to a Gaussian shape, namely,
\begin{equation}
\Psi({\mathbf r},t) \simeq \frac{e^{-\frac{|{\mathbf r}|^2}{4Dt}}}{\left(4\pi D t \right)^{d/2} },
\label{boltzmann_diffusion}
\end{equation}
the quantity $D=v/(d \sigma)$ playing the role of a diffusion coefficient~\cite{cercignani}. However, Eq.~\eqref{boltzmann_diffusion} is approximately valid for $r \sigma \gg 1$, and can not capture the particle evolution at early times, nor the finite-speed propagation effects. Indeed, diffusion implicitly assumes a non-vanishing probability of finding the particles at arbitrary distance from the source. Deviations from the limit Gaussian behavior are well known, e.g., for neutron~\cite{cercignani} as well electron transport~\cite{rev_elec_1}. 

In the following, we outline the relation between $\Psi({\mathbf r},t)$ and the underlying exponential flight process.

\subsection{The free propagator without absorptions}

Consider a $d$-dimensional setup. A particle, originally located at position ${\mathbf r_0}$ in a given domain, travels along straight lines at constant speed $v$, until it collides with the medium. The position of a particle at the $n$-th collision can be expressed as a random walk ${\mathbf r}_n={\mathbf r}_0+\sum_{i=1}^n {\mathbf r}_i$, i.e., a sum of random variables ${\mathbf r}_i$. The flight lengths $\ell=|{\mathbf r}_i-{\mathbf r}_{i-1}|$ are assumedly identically distributed and obey an exponential probability density
\begin{equation}
\varphi(\ell)= \sigma_t e^{-\ell \sigma_t},
\label{exp_flight_length}
\end{equation} 
with $\sigma_t>0$. The exponential law in Eq.~\eqref{exp_flight_length} stems from assuming a uniform distribution of the scattering centers in the traversed medium. Heterogeneous materials, such as complex fluids, would generally lead to clustered scattering centers, obeying to, e.g., negative binomial distributions, and in turn to non-exponential flight lengths~\cite{anisotropy}. However, we will focus our attention on homogeneous media.

At each collision, the particle randomly changes its direction according to the scattering kernel $k({\mathbf \omega'},{\mathbf \omega})$. For the sake of simplicity, we assume here that the scattering is isotropic, so that $k({\mathbf \omega'},{\mathbf \omega})$ has a uniform distribution, independent of the incident direction ${\mathbf \omega'}$.

Once a flight length has been sampled from $\varphi(\ell)$, the new direction ${\mathbf \omega}$ is therefore uniformly distributed on the $d$-sphere of surface $\ell^{d-1}\Omega_d$. Therefore, by virtue of the apparent spherical symmetry, the transition kernel, i.e., the probability density of performing a displacement from ${\mathbf r}_{i-1}$ to ${\mathbf r}_i$, depends only on $\ell=|{\mathbf r}_i-{\mathbf r}_{i-1}|$, and reads
\begin{equation}
\pi_d(\ell)=\frac{\varphi(\ell)}{\ell^{d-1}\Omega_d}.
\end{equation}

We initially neglect absorptions, so that $\sigma_t=\sigma$: in one-speed transport, this condition can be either seen as the particles been scattered, or equivalently being absorbed and then re-emitted (with the same speed) at each collision. This latter interpretation would correspond, e.g., to a {\em criticality} condition in multiplicative systems for neutron transport.

We define then the free propagator $\Psi({\mathbf r}|n)$ as the probability density of finding a particle at position ${\mathbf r}$ at the $n$-th collision, for an infinite medium, i.e., in absence of boundaries. We adopt here the convention that the particle position and direction refer to the physical properties before entering the collision; for instance, the index $n=1$ refers to uncollided particles, i.e., particles coming from the source and entering their first collision.

Assuming that all the particles are isotropically emitted at ${\mathbf r_0}={\mathbf 0}$, the particle density $\Psi({\mathbf r}|n)$ must depend only on the variable $r=|{\mathbf r}|$, because of the spherical symmetry. On the basis of the properties exposed above, the particle propagation as a function of the number of collisions is a Markovian process in the variable ${\mathbf r_n}$, where for each collision $i=1,...,n$ the new propagator is given by the convolution integral
\begin{equation}
\Psi(r|i)=\int \pi_d(|{\mathbf r}-{\mathbf r'}|) \Psi(r'|i-1) d{\mathbf r'},
\label{convolution_int}
\end{equation}
with initial condition $\Psi(r|0)=\delta(r)$. In particular, by direct integration we immediately get the uncollided propagator
\begin{equation}
\Psi(r|1)=\pi_d(r).
\end{equation}

It is convenient to introduce the $d$-dimensional Fourier transform of spherical-symmetrical functions, as in the subsequent analysis this will allow easier deriving of the properties of the exponential flights. Denoting by $z$ the transformed variable with respect to $r$, for any spherical-symmetrical function $f(r)$ we have the following transform and anti-transform pair $f(z)={\cal F}_d\left\lbrace f(r)\right\rbrace $ and $f(r)={\cal F}_d^{-1}\left\lbrace f(z)\right\rbrace $~\cite{dutka}
\begin{eqnarray}
f(z)=z^{1-d/2}\left(2\pi \right)^{d/2}\int_0^{+\infty} r^{d/2} J_{d/2-1}(zr) f(r) dr \nonumber \\
f(r)=r^{1-d/2}\left(2\pi \right)^{-d/2}\int_0^{+\infty} z^{d/2} J_{d/2-1}(rz) f(z) dz,
\label{fourier_definition}
\end{eqnarray}
where $J_\nu()$ is the modified Bessel function of the first kind, with index $\nu$~\cite{abramowitz}. It is apparent from Eqs.~\eqref{fourier_definition} that the dimension $d$ of the system plays a fundamental role, in that it affects both the transition kernel and the Fourier transform kernel itself.

\begin{figure}[t]
   \centerline{ \epsfclipon \epsfxsize=9.0cm
\epsfbox{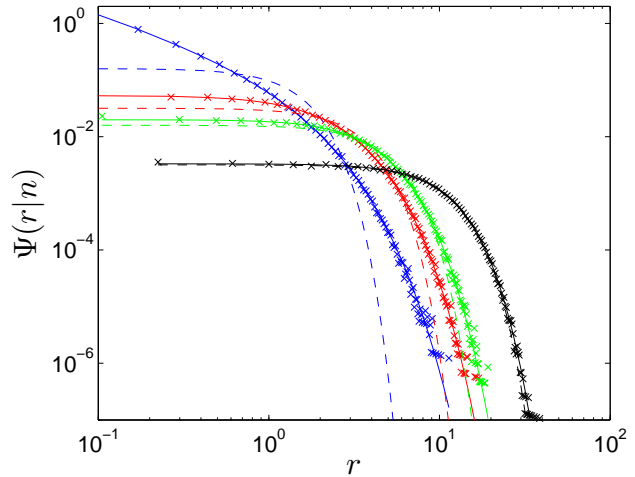} }
   \caption{The free propagator $\Psi(r|n)$ for $d=2$, with $n=1$ (blue), $5$ (red), $10$ (green), and $50$ (black). Monte Carlo simulation results are displayed in symbols. The solid line is the theoretical result, Eq.~\eqref{2d_propagator}. The dashed line is the diffusion limit, Eq.~\eqref{diff_eq_n}.}
   \label{figpn_d2}
\end{figure}

The convolution integral in Eq.~\eqref{convolution_int} in Fourier space gives the algebraic relation
\begin{equation}
\Psi(z|i)=\pi_d(z) \Psi(z|i-1),
\label{convolution_prod}
\end{equation}
$i \ge 1$, with initial condition $\Psi(z|0)=1$. By recursion, it follows that in the transformed space
\begin{equation}
\Psi(z|n)=\pi_d(z)^n.
\end{equation}
It turns out that the Fourier transform of the transition kernel $\pi_d(z)$ can be explicitly performed in arbitrary dimension, and gives
\begin{equation}
\pi_d(z)=~_2F_1\left(\frac{1}{2},1,\frac{d}{2};-\frac{z^2}{\sigma^2}  \right),
\end{equation}
where $_2F_1$ is the Gauss hypergeometric function~\cite{abramowitz}. Hence, we finally obtain
\begin{equation}
\Psi(z|n)=\left[_2F_1\left(\frac{1}{2},1,\frac{d}{2};-\frac{z^2}{\sigma^2} \right)\right]^n.
\label{p_z_n}
\end{equation}
The quantity $\pi_d(z)$ is positive for $d \ge 1$; moreover, $\pi_d(z=0)=1$, which ensures normalization and positivity of the propagator. In Fig.~\ref{fig1} we visually represent the effects of dimension and number of collisions on the shape of $\Psi(z|n)$. Remark in particular that the spread of $\Psi(z|n)$ increases with $d$, for a given $n$. On the contrary, $\Psi(z|n)$ becomes more peaked around the origin with growing $n$, for a given $d$.

Formally, performing the inverse Fourier transform of Eq.~\eqref{p_z_n} gives the propagator $\Psi(r|n)={\cal F}_d^{-1}\left\lbrace \Psi(z|n)\right\rbrace$ for an arbitrary $d$-dimensional setup. Actually, in some cases this task turns out to be non-trivial. Nonetheless, even in absence of an explicit functional form for the propagator, information can be extracted by resorting to the Tauberian theorems. In particular, the expansion of $\Psi(z|n)$ for $ z/\sigma \ll 1$ gives the behavior of $\Psi(r|n)$ for $r \sigma \gg 1$, i.e., far from the source, in the so-called {\em diffusion limit}~\cite{feller, klafter}; viceversa, the expansion of $\Psi(z|n)$ for $z/\sigma \gg 1$ gives the behavior of $\Psi(r|n)$ for $r\sigma \ll 1$, i.e., close to the source. We recall that $_2F_1$ is defined through the series~\cite{abramowitz}
\begin{equation}
_2F_1\left(\frac{1}{2},1,\frac{d}{2};-\frac{z^2}{\sigma^2} \right)=\sum_{k=0}^{\infty} \frac{\Gamma\left( \frac{d}{2}\right) \Gamma\left(\frac{1}{2}+k \right) }{\sqrt{\pi}\Gamma\left(\frac{d}{2}+k\right) }\left( i\frac{z}{\sigma}\right) ^{2k}.
\label{hypergeom_expansion}
\end{equation}
At the leading order for $z/\sigma \to 0$ we therefore have
\begin{equation}
\pi_d(z) \simeq 1-\frac{1}{d}\left( \frac{z}{\sigma}\right)^2 + ...
\label{gauss_exp}
\end{equation}
Observe that Eq.~\eqref{gauss_exp} can be viewed as the expansion of an exponential function. Then, the inverse Fourier transform gives the Gaussian shape
\begin{equation}
\Psi(r|n) \simeq \frac{e^{-\frac{r^2 }{4n{\cal D}}}}{\left( 4\pi{\cal D}n\right)^{d/2}},
\label{diff_eq_n}
\end{equation}
which is valid for $r\sigma \gg 1$, ${\cal D}=1/(d\sigma^2)$ playing the role of a diffusion coefficient. This stems from the exponential flights having finite-variance increments, $\langle \ell^2 \rangle < +\infty$, so that their probability density $\Psi(r|n)$ falls in the basin of attraction of the Central Limit Theorem~\cite{feller}. We mention that clustered scattering centers, with non-exponential flight lengths, may lead to non-Gaussian limiting statistics~\cite{anisotropy}. Remark the close analogy between Eqs.~\eqref{boltzmann_diffusion} and~\eqref{diff_eq_n}: in particular, ${\cal D}$ and $D$ differ by a factor $\sigma v$, which roughly speaking represents the average number of collisions per unit time.

Moreover, at the leading order for $z/\sigma \to \infty$ we have the expansion
\begin{equation}
\pi_d(z) \simeq \frac{\sqrt{\pi}\Gamma\left(\frac{d}{2}\right)}{\Gamma\left(\frac{d-1}{2}\right)}\left(\frac{\sigma}{z} \right)+(2-d)\left(\frac{\sigma}{z} \right)^{2}+...,
\label{inf_exp}
\end{equation}
where the first term vanishes for $d=1$. By inverse Fourier transforming, we have for $r\sigma \ll 1$
\begin{equation}
\Psi(r|n) \simeq c_{1}^{n,d} + c_{2}^{n,d} (r\sigma)^{n-d} ,
\label{source_eq_n}
\end{equation}
when $d>1$, and $\Psi(r|n) \simeq c_{1}^{n,1} + c_{2}^{n,1} (r\sigma)^{2n-1}$ when $d=1$. Here $c_{1}^{n,d}$ and $c_{2}^{n,d}$ are constants depending on $n$ and $d$. It can be shown that the divergence at the origin in $\Psi(r|n)$ due to the Dirac delta source disappears after $n > d$ collisions for $d>1$, and $n \ge 1$ for $d=1$.

\begin{figure}[t]
   \centerline{ \epsfclipon \epsfxsize=9.0cm
\epsfbox{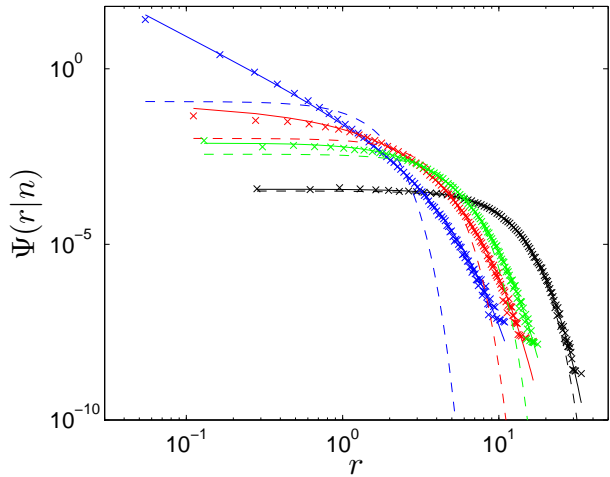} }
   \caption{The free propagator $\Psi(r|n)$ for $d=3$, with $n=1$ (blue), $5$ (red), $10$ (green), and $50$ (black). Monte Carlo simulation results are displayed in symbols. The dashed line is the diffusion limit, Eq.~\eqref{diff_3d}. The solid line is the approximate propagator, Eq.~\eqref{approximate_3d}.}
   \label{figpn_d3}
\end{figure}

\subsection{Relation between collision number and time}

Assume again that $\sigma_t=\sigma$, i.e., that there are no absorptions. The free propagator $\Psi(r|n)$ gives information on the position of a transported particle at the moment of entering the $n$-th scattering collision. The link between the travelled distance, the flight time and the number of collisions is provided by the speed $v$. Indeed, once a flight of length $\ell$ between any two collisions has been sampled from $\varphi(\ell)$, the flight time must satisfy $t_\ell= t_{i}-t_{i-1}=\ell/v$.  Hence, the transition kernel, i.e., the probability density of performing a displacement from ${\mathbf r}_{i-1}$ at $t_{i-1}$ to ${\mathbf r}_i$ at $t_{i}$, will be given by
\begin{equation}
\pi_d(\ell,t_\ell)=\pi_d(\ell)\delta\left( t_\ell-\frac{\ell}{v} \right).
\end{equation}
It follows that inter-collision times are exponentially distributed
\begin{equation}
w(t_\ell)=\int \pi_d(\ell,t_\ell) \Omega_d \ell^{d-1}d\ell=\frac{e^{-t_\ell/\tau}}{\tau},
\end{equation}
where $\tau=1/(\sigma v)$ represents the average time between collisions.

We define the propagator $\Psi(r,t|n)$ as the probability density of finding a particle at position ${\bf r}$ at time $t$ at the $n$-th collision. From the Markov property of the process, at each collision $i=1,...,n$ we have
\begin{equation}
\Psi(r,t|i)=\int d{\mathbf r'} \int d{t'}\pi_d(|{\mathbf r}-{\mathbf r'}|,t-t') \Psi(r',t'|i-1),
\label{convolution_int_time}
\end{equation}
with initial condition $\Psi(r,t|0)=\delta(r)\delta(t/\tau)=\tau{\cal S}(r,t)$. In particular, by direct integration we immediately get the uncollided propagator
\begin{equation}
\Psi(r,t|1)=\pi_d(r,t)=\tau \Psi(r|1)\delta\left( t-\frac{r}{v}\right).
\label{prt1}
\end{equation}

\begin{figure}[t]
   \centerline{ \epsfclipon \epsfxsize=9.0cm
\epsfbox{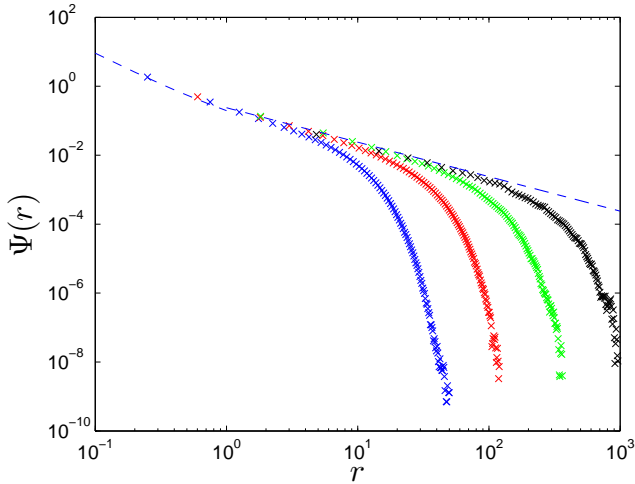} }
   \caption{The collision density $\Psi(r)$ for $d=3$, with increasing number of summed collisions, $N=10^2$ (blue), $10^3$ (red), $10^4$ (green), and $10^5$ (black). The dashed lines are the asymptotic limits close (Eq.~\eqref{close_source_3d}) and far (Eq.~\eqref{far_source_3d}) from the source.}
   \label{figpsi_3d}
\end{figure}

We denote the Laplace transform of a function $g(t)$ by its argument, i.e.,
\begin{equation}
g(s)={\cal L}\left\lbrace g(t)\right\rbrace =\int_0^{+\infty}e^{-st}g(t)dt.
\end{equation}
Then, from the double convolution integral in Eq.~\eqref{convolution_int_time} we have the algebraic product in the Fourier and Laplace space
\begin{equation}
\Psi(z,s|i)=\pi_d(z,s) \Psi(z,s|i-1),
\label{convolution_transf}
\end{equation}
$i \ge 1$, with $\Psi(z,s|0)=\tau$. By recursion, it follows that in the transformed space we have
\begin{equation}
\Psi(z,s|n)=\tau\pi_d(z,s)^n.
\end{equation}
It turns out that the Fourier and Laplace transform of the transition kernel $\pi_d(z,s)$ can be explicitly performed in arbitrary dimension, and gives
\begin{equation}
\pi_d(z,s)=\frac{_2F_1\left(\frac{1}{2},1,\frac{d}{2};- \frac{z^2}{\zeta^2}\right)}{1+s\tau},
\end{equation}
where $\zeta(s)=\sigma(1+s\tau)$. Hence, we finally obtain
\begin{equation}
\Psi(z,s|n)=\tau\left[ \frac{_2F_1\left(\frac{1}{2},1,\frac{d}{2};-\frac{z^2}{\zeta^2}\right)}{1+s\tau}\right] ^n.
\label{propagator_f_l}
\end{equation}

Moreover, the following relation follows: $\Psi(z,s=0|n)=\tau \Psi(z|n)$, so that
\begin{equation}
\Psi(r|n) = \frac{1}{\tau}\int_0^{+\infty}\Psi(r,t|n)dt,
\label{propagator_partial}
\end{equation}
i.e., $\Psi(r|n)$ can be intepreted as the time average of $\Psi(r,t|n)$. Finally, the propagator $\Psi(r,t)$ will be given by the sum of the contributions $\Psi(r,t|n)$ at each collision, namely
\begin{equation}
\Psi(r,t)=\sum_{n=1}^{\infty}\Psi(r,t|n).
\label{psi_r_t}
\end{equation}
Taking the Fourier and Laplace transform of Eq.~\eqref{psi_r_t}, we then have
\begin{equation}
\Psi(z,s)=\sum_{n=1}^{\infty}\Psi(z,s|n)=\tau \frac{\pi_d(z,s)}{1-\pi_d(z,s)},
\end{equation}
with $\Psi(r,t)={\cal L}^{-1}{\cal F}_d^{-1}\left\lbrace \Psi(z,s) \right\rbrace$.

\subsection{Absorptions}

In presence of absorptions ($\sigma_a>0$), the propagator $\Psi_a(r,t)$ can be obtained from Eq.~\eqref{boltzmann_absorption} by integrating over directions. This relation holds true at each collision, so that we have
\begin{equation}
\Psi_a(r,t|n) = \Psi(r,t|n) e^{-t/\tau_a},
\end{equation}
where $\tau_a=1/(\sigma_a v)$. Hence, from Eq.~\eqref{propagator_partial}, by replacing $\sigma$ with $\sigma_t$ we get for the propagator $\Psi_a(z|n)$
\begin{equation}
\Psi_a(z|n) = \frac{1}{\tau_t} \int_0^{+\infty}\Psi_a(z,t|n)dt=\frac{1}{\tau_t}\Psi(z,s=\frac{1}{\tau_a}|n),
\end{equation}
where $\tau_t=1/(\sigma_t v)$ represents the average flight time between any two collisions. Now, observe that from Eq.~\eqref{propagator_f_l} we have
\begin{equation}
\Psi\left( z,s=\frac{1}{\tau_a}\vert n\right)  = \tau \Psi\left( z \frac{\sigma}{\sigma_t}\vert n\right)\left(  \frac{\sigma}{\sigma_t}\right)^n ,
\end{equation}
where the quantity $p=\sigma/\sigma_t$, $0<p<1$, can be interpreted as the the probability of being scattered, i.e., {\em not} being absorbed, at any given collision. Then, by remarking that $\tau/\tau_t=1/p$, it follows that
\begin{equation}
\Psi_a(z|n)=\Psi\left(p z |n\right)  p^{n-1}.
\label{propagator_absorption}
\end{equation}
The propagator in Eq.~\eqref{propagator_absorption} is then given by the product of the free propagator, with the total cross section $\sigma_t$ replacing the scattering cross section $\sigma$, times the probability of having survived up to entering the $n$-th collision. Remark that the total cross section and the non-absorption probability are related by $\sigma = p \sigma_t$. When the absorption length is infinite, $\sigma_a \to 0$, $p \to 1$ and we recover the free propagator for pure scattering, with $\sigma_t = \sigma$.

\subsection{Boundary conditions}

So far, we have assumed that the medium where particles are transported has an infinite extension, hence the name free propagator. More realistically, we might consider finite-extension media with volume ${\cal V}$ enclosing the source, so that boundary conditions come into play and affect the nature of the propagator. Several boundary conditions can be conceived according to the specific physical system, among which the most common are reflecting and leakage. This issue has been extensively examined, e.g., for radiation shielding in Reactor Physics~\cite{bell, wigner, cercignani} and for electron motion in semiconductors~\cite{elec_leak_1, elec_leak_2}. Here we will focus on leakage boundary conditions, where particles are considered lost as soon as their trajectory has traversed the outer boundary $\partial {\cal V}$ of the domain. While the volume ${\cal V}$ is in principle totally arbitrary, in the subsequent calculations for the sake of convenience we will assume that ${\cal V}={\cal V}(R)$ is a sphere of radius $R$ centered in ${\bf r_0}={\bf 0}$.

From the point of view of the propagator, leakages can be taken into account by assuming that the population density $\Psi(r|n)$ at any $n$ vanishes at the so-called extrapolation length $r_e$, i.e., $\Psi(r=r_e|n)=0$~\cite{bell, wigner, cercignani}. Because trajectory do not terminate at the boundary, but rather at the first collision occurred outside the volume, the extrapolation length is expected to be larger than the physical boundary of ${\cal V}(R)$ and can be determined from solving the so-called Milne's problem associated to the volume~\cite{placzek, freund}. In general, $r_e$ is of the kind $r_e=R\left[ 1+u_d/(R\sigma)\right] $, the dimensionless constant $u_d>0$ depending on the dimension of the system~\cite{freund}. When the scattering length is much smaller than the typical size of the volume, i.e., $\sigma R \gg 1$, the extrapolation length coincides with the physical boundary, $r_e \to R$. This means that the inter-collision length is so small as compared to the total travelled distance that the first collision outside the domain will actually be very close to the last collision inside the domain, which corresponds to $\Psi(r|n)$ vanishing at $R$~\cite{feller, cercignani}.

\section{Spatial moments of the propagator}
\label{space_moments}

The moments of the propagator provide an estimate of the spatial evolution of the particles ensemble, as a function of the number of collisions or time. Due to the supposed spherical symmetry, we expect all the odd moments to vanish. We define the $m$-th moment of $f(r)$ over the spherical shell $r^{d-1} \Omega_d dr$ as
\begin{eqnarray}
\langle r^m \rangle = \Omega_d \int_0^{+\infty} r^{m+d-1}f(r) dr.
\label{mom_int}
\end{eqnarray}
Performing the integral in Eq.~\eqref{mom_int} would require explicitly knowing $f(r)$. However, as shown in the Appendix~\ref{appendix a}, the $m$-th spatial moment $f(r)$ of can be expressed as a function of the $m$-th derivative of $f(z)$ with respect to $z$
\begin{equation}
\langle r^m \rangle=\frac{\sqrt{\pi}\Gamma\left( \frac{d+m}{2}\right) }{\Gamma\left( \frac{d}{2}\right)\Gamma\left( \frac{1+m}{2}\right)} \frac{\partial^m}{\partial (iz)^m}  \left[ f(z)\right]_{z=0}
\label{spatial_moments}
\end{equation}
when $m$ is even, and $\langle r^m \rangle=0$ otherwise. By setting $f(z)=\Psi(pz|n)p^{n-1}$ in Eq.~\eqref{spatial_moments} we then have $\langle r^m \rangle(n)$ as a function of the number of collisions. Analogously, by setting $f(z)=\Psi_a(z,s)=\Psi(z,s+1/\tau_a)$ we get $\langle r^m \rangle(s)$, which gives the evolution as a function of time, upon inverse Laplace transforming. In particular, for the spread $m=2$ of the propagator with absorptions we get
\begin{equation}
\langle r^2 \rangle(n)=\frac{2}{\sigma_t^2}p^{n-1}n,
\end{equation}
and
\begin{equation}
\langle r^2 \rangle(t)=\frac{2}{\sigma^2}\left[e^{-t/\tau}-1 + \frac{t}{\tau}\right] e^{-t/\tau_a}.
\end{equation}
In absence of absorption, $\tau_a \to \infty$ and $p \to 1$, the particle spread $\langle r^2 \rangle(n)$ is linear with respect to $n$. On the contrary, $\langle r^2 \rangle(t)$ has a ballistic behavior ($\propto t^2$) at earlier times (where transport is dominated by velocity), and a diffusive behavior ($\propto t$) at later times (where transport is dominated by scattering). The transition between the two regimes is imposed by the time scale $\tau$. A remarkable feature is that in either case the spread does not explicitly depend on the dimension $d$. Intuitively, this can be understood by considering that (independently of the dimension $d$ of the embedding setup) the vectors ${\bf r}_i$ and ${\bf r}_{i+1}$ define a plane with random orientation, so that space is explored by plane surfaces at each collision. This is in analogy, e.g., with the behavior of $d$-dimensional Brownian Motion~\cite{feller}.

\section{Collision density and collision statistics}
\label{coll_density_sec}

In many physical problems, one is interested in assessing the statistics of the time $t_{\cal V}$ or the collision number $n_{\cal V}$ spent inside a given domain ${\cal V}$. A prominent role in characterizing a physical system is played in particular by the mean residence time $\langle t_{\cal V} \rangle$ and the mean collision number $\langle n_{\cal V} \rangle$~\cite{redner}. In Reactor Physics, for instance, the average number of neutron collisions within a region would be related to such issues as the nuclear heating or nuclear damage in fissile as well as structural materials~\cite{bell, wigner}. We introduce the collision density $\Psi(r)$~\cite{case}, which is defined as
\begin{equation}
\Psi(r)=\lim_{N \to \infty} \sum_{n=1}^N \Psi(r|n).
\label{collision_density}
\end{equation}
From Eq.~\eqref{psi_r_t}, it follows that $\Psi(r)$ can be equivalently obtained from the propagator $\Psi(r,t)$ as
\begin{equation}
\Psi(r)= \lim_{T \to \infty} \frac{1}{\tau}\int_0^{T}\Psi(r,t)dt.
\end{equation}
For an infinite `observation time' $T$, the mean residence time $\langle t_{\cal V} \rangle$ within ${\cal V}$ can be expressed in terms of the collision density $\Psi(r)$ in the same domain by slightly adapting an argument due to Kac~\cite{kac, berezhkovskii}
\begin{equation}
\frac{\langle t_{\cal V} \rangle}{\tau} = \int_{{\cal V}}d {\mathbf r}_1 \Psi(r_1),
\label{time_moments}
\end{equation}
where $r_1=|{\mathbf r}_1-{\mathbf r}_{0}|$.

As for the collision number, the probability of performing $n_{\cal V}$ collisions in ${\cal V}$ is given by
\begin{equation}
{\cal P}(n_{{\cal V}})=\int_{{\cal V}}d {\mathbf r} \Psi(r|n_{{\cal V}})-\int_{{\cal V}}d {\mathbf r} \Psi(r|n_{{\cal V}}+1),
\end{equation}
hence the moments
\begin{equation}
\langle n_{\cal V}^m \rangle = \sum_{n_{{\cal V}}=1}^{+\infty} n_{{\cal V}}^m {\cal P}(n_{{\cal V}}).
\label{n_poisson}
\end{equation}
From the definition of $\Psi(r)$, it follows immediately that
\begin{equation}
\langle n_{\cal V} \rangle = \frac{\langle t_{\cal V} \rangle}{\tau},
\label{average_n_t}
\end{equation}
i.e., the integral of the collision density over a volume ${\cal V}$ gives the mean number of collisions within that domain, hence the name given to $\Psi(r)$.

Remark that both $\langle t_{\cal V} \rangle$ and $\langle n_{\cal V} \rangle$ depend on the boundary conditions imposed on $\partial {\cal V}$, which in turn affect the functional form of the propagator, and then $\Psi(r)$. Using a free propagator corresponds to defining a fictitious volume ${\cal V}$, whose boundaries $\partial {\cal V}$ are `transparent', so that particles can indefinitely cross $\partial {\cal V}$ back an forth. On the contrary, the use of leakage boundary conditions leads to the formulation of first-passage problems, i.e., determining the distribution of the time, or collision number, required to first reach the boundary. Consequently, $\langle t_{\cal V} \rangle$ is preferentially called mean first-passage time~\cite{redner}.

It can be shown that higher order moments of $n_{\cal V}$ can be obtained by recursion, in terms of Kac integrals analogous to those in Eq.~\eqref{time_moments}. A detailed treatment is beyond the scope of the present paper and is discussed in~\cite{zoia_arxive}.

The previous discussion shows that $\Psi(r)$ is key in determining residence times and collisions statistics. After formally carrying out the summation in Eq.~\eqref{collision_density}, we have $\Psi(z)=\pi_d(z)/[1-\pi_d(z)]$, so that $\Psi(r)$ for a free propagator in absence of absorption is defined in terms of the following Fourier integral 
\begin{equation}
\Psi(r)=\frac{r^{1-d/2}}{\left(2\pi \right)^{d/2}}\int_0^{+\infty} z^{d/2} J_{d/2-1}(rz) \frac{\pi_d(z)}{1-\pi_d(z)} dz,
\label{coll_density_def}
\end{equation}
whose convergence depends on the dimension $d$ of the system. It turns out that convergence is ensured for $d>2$, which means that for $1d$ and $2d$ finite-size domains with transparent boundaries $\langle n_{\cal V} \rangle$ and $\langle t_{\cal V} \rangle$ diverge as $N \to \infty$. This result is in analogy with P\'olya's theorem, which states that random walks on Euclidean lattices are recurrent for $d\le 2$~\cite{feller}. As shown in the following, we can nonetheless provide an estimate of such divergence as a function of $N$, i.e., single out a singular term from a functional form. For finite domains with leakage boundary conditions and/or absorptions ($\sigma_a>0$), $\Psi(r)$ is defined also for $1d$ and $2d$ systems. For $d>2$, Tauberian theorems show that the asymptotic behavior of Eq.~\eqref{coll_density_def} is given by
\begin{equation}
\Psi(r) \simeq \frac{\Gamma\left(\frac{d}{2} \right) }{2 \pi^{d/2}}(r\sigma)^{2-d}
\end{equation}
for large $r$, and
\begin{equation}
\Psi(r) \simeq \frac{\Gamma\left(\frac{d}{2} \right) }{2 \pi^{d/2}}(r\sigma)^{1-d}
\end{equation}
close to the source.

\section{Analysis of $d$-dimensional setups}
\label{dimensional_analysis}

In the following, we detail the results pertaining to specific values of $d$. We choose $1/\sigma_t$ as length scale and we work with dimensionless spatial variables $r=r\sigma_t$. Remark that in absence of absorption the length scale is $1/\sigma$, since $p=1$.

\subsection{One-dimensional setup, $d=1$}
\label{onedimension}

The case $d=1$ allows illustrating the general structure of the calculations. One potential application of this framework could be provided by nanowires or carbon nanotubes (almost $1d$ systems) in electron transport~\cite{jacoboni_book}. The transition kernel is
\begin{equation}
\pi_1(\ell)=\frac{e^{-\ell}}{2},
\end{equation}
whose Fourier transform is
\begin{equation}
\pi_1(z)=\frac{1}{1+z^2}.
\end{equation}
Starting from $\Psi(z,0)=1$, the free propagator $\Psi(r|n)$ can be explicitly obtained by performing the inverse Fourier transform of $\Psi(z|n)=\pi_1(z)^n$, and reads
\begin{equation}
\Psi(r|n)=\frac{2^{\frac{1}{2}-n}r^{-\frac{1}{2}+n}K_{-\frac{1}{2}+n}\left( r \right)  }{\sqrt{\pi}\Gamma\left( n\right) },
\label{1d_propagator}
\end{equation}
where $K_\nu()$ is the modified Bessel function of the second kind, with index $\nu$~\cite{abramowitz}. The same formula has been recently derived, e.g., in~\cite{lecaer} as a particular case of a broader class of random flights. In Fig.~\ref{figpn_d1} we provide a comparison between Monte Carlo simulation results (symbols), the analytical formula Eq.~\eqref{1d_propagator} (solid lines), and the diffusion limit, Eq.~\eqref{diff_eq_n} (dashed lines), for different values of $n$. Remark in particular that the diffusion limit is not accurate for small $n$, and becomes progressively closer to the exact result for increasing $n$, as expected. At intermediate $n$, the tails of the propagator~\eqref{1d_propagator} are always fatter than those predicted by the diffusion approximation.

In a $1d$ setup, the collision density $\Psi(r)$ for the free propagator diverges. Nonetheless, it is possible to single out the divergence as follows
\begin{equation}
\Psi(z)=\lim_{N \to \infty}\sum_{n=1}^N \Psi(z|n) =\lim_{N \to \infty}\frac{1-\left(1+z^2  \right)^{-N} }{z^2 }.
\end{equation}
For fixed $N$, the inverse transform can be explicitly performed in terms of hypergeometric functions. Retaining the non-vanishing terms for large $N$, we have
\begin{equation}
\Psi(r) \simeq  \frac{\Gamma\left(\frac{1}{2}+N \right) }{\Gamma\left(N \right)\sqrt{\pi}}-\frac{r}{2}  \simeq  \frac{\sqrt{N}}{\sqrt{\pi}}-\frac{r}{2},
\end{equation}
which is composed of a term diverging with $\sqrt{N}$ (not depending on $r$), and a functional part which is linear in $r$ (not depending on $N$).

For the propagator with absorptions, from Eq.~\eqref{propagator_absorption} we have
\begin{equation}
\Psi_a(z)=\lim_{N \to \infty} \sum_{n=1}^N \Psi\left( z|n\right) p^{n-1}=\frac{1}{1-p+z^2}.
\end{equation}
Then, performing the inverse Fourier transform, we get
\begin{equation}
\Psi_a(r)=\frac{e^{-\sqrt{1-p}r}}{2\sqrt{1-p}}.
\label{psia_with_abs}
\end{equation}
Remark that Eq.~\eqref{psia_with_abs} has been derived, e.g., in~\cite{wing} by solving the stationary Boltzmann equation in $1d$. When $p \to 0$, the particles are almost surely absorbed at the first collision, and we have the expansion
\begin{equation}
\Psi_a(r) \simeq \frac{e^{-r}}{2}\left(1+\frac{p}{2}+\frac{pr}{2} \right)+...,
\end{equation}
so that at the first order the collision density has the same functional form as the uncollided propagator. At the opposite, when $p \to 1$ the particles are almost surely always scattered ($\sigma_t \to \sigma$), and we have the expansion
\begin{equation}
\Psi_a(r) \simeq \frac{1}{2\sqrt{1-p}}-\frac{r}{2} +...,
\end{equation}
and $\Psi_a(r)$ diverges as the collision density associated to the free propagator, as expected.

The case of leakage boundary conditions can be dealt with by imposing that the propagator $\Psi(r|n)$ must vanish for any $n$ at the extrapolated boundary $r_e$. For $d=1$, the extrapolated length is given by $r_e=R\left[ 1+u_d/R\right] $ with $u_1=1$~\cite{freund}, i.e., the propagator must vanish at one scattering length outside the physical border $R$ of the domain. By resorting to the method of images~\cite{feller}, which allows solving for the propagator in presence of boundaries in terms of the propagator in absence of boundaries, we therefore have for the collision density
\begin{equation}
\Psi_R(r) = \frac{r_e-r}{2},
\label{coll_dens_1d_leak}
\end{equation}
for $r \le R$, and $\Psi(r)=0$ elsewhere.

The moments of the residence time (or, equivalently, of the number of collisions) within a sphere of radius $R$ can be explicitly computed based on Eq.~\eqref{average_n_t}. The moments associated to the free propagator clearly diverge. Here we separately analyze the propagator with absorptions (in an infinite domain) and the propagator with leakages at the boundary $r=r_e \ge R$ (without absorption). For the case of absorptions, the average residence time within a (fictitious) sphere of radius $R$ reads
\begin{equation}
\langle t_R \rangle = \frac{1-e^{-\sqrt{1-p}R}}{1-p} \tau_t,
\end{equation}
assuming that particles can cross the boundaries of the sphere an infinite number of times. When the radius of the sphere is large as compared to the typical particle displacement, we have $\langle t_R \rangle \simeq 1/(1-p) \tau_t$, which gives $\langle t_R \rangle \simeq \tau_t$ when $p \to 0$, and diverges for $p \to 1$. For the case of leakages, the mean first-passage time reads
\begin{equation}
\langle t_R \rangle = \frac{R \left(2r_e-R \right) }{2} \tau.
\end{equation}
When the radius of the sphere is large as compared to the typical particle displacement, $r_e \to R$, and we have $\langle t_R \rangle \simeq R^2\tau/2 $.

\subsection{Two-dimensional setup, $d=2$}
\label{twodimension}

The case $d=2$ has a key interest in assessing, e.g., the dynamics of chemical and biological species on surfaces~\cite{bacteria}. Moreover, it concerns also quantum wells and inversion layers in electron transport~\cite{jacoboni_book}. The transition kernel reads
\begin{equation}
\pi_2(\ell)=\frac{ e^{-\ell}}{2\pi \ell},
\end{equation}
whose Fourier transform is
\begin{equation}
\pi_2(z)=\frac{1}{\sqrt{1+ z^2}}.
\end{equation}
Starting from $\Psi(z,0)=1$, the free propagator $\Psi(r|n)$ can be explicitly obtained by performing the inverse Fourier transform of $\Psi(z|n)=\pi_2(z)^n$, and reads
\begin{equation}
\Psi(r|n)=\frac{ 2^{-\frac{n}{2}}r^{-1+\frac{n}{2}}K_{-1+\frac{n}{2}}\left(r  \right)  }{\pi\Gamma\left( \frac{n}{2}\right) }.
\label{2d_propagator}
\end{equation}
This result was previously established by~\cite{stadje}, and has later appeared in, e.g.,~\cite{connolly, lecaer}. In Fig.~\ref{figpn_d2} we compare the Monte Carlo simulation results (symbols) with the theoretical formula in Eq.~\eqref{2d_propagator}, for different values of $n$. The diffusion limit, Eq.~\eqref{diff_eq_n}, is also shown in dashed lines. Remark that the diffusion limit is not accurate for small $n$, and becomes progressively closer to the exact result for increasing $n$. At intermediate $n$, the tails of the propagator~\eqref{2d_propagator} are always fatter than those predicted by the diffusion approximation.

In a $2d$ setup, the collision density $\Psi(r)$ diverges. Nonetheless, analogously as done for the $1d$ case, it is possible to single out the divergence as follows
\begin{equation}
\Psi(z)=\lim_{N \to \infty} \sum_{n=1}^N \Psi(z|n)=\lim_{N \to \infty}\frac{1-\left(1+z^2  \right)^{-N/2} }{\sqrt{1+z^2} -1}.
\end{equation}
For fixed $N$, the inverse transform can be explicitly performed. Details of the rather cumbersome calculations are left to the Appendix~\ref{appendix b}. Retaining the non-vanishing terms for large $N$, we have
\begin{eqnarray}
\Psi(r) \simeq \frac{\log\left(N\right)}{2\pi} +  \frac{ e^{-r}}{2 \pi r} + \frac{{\text Ei}\left(- r\right) -2\log\left(r \right)}{2\pi },
\label{coll_dens_free_2d}
\end{eqnarray}
where ${\text Ei}$ is the exponential integral function~\cite{abramowitz}. To the authors' best knowledge, the formula for the collision density $\Psi(r)$ has not appeared before, and might then provide a useful tool for describing the migration of species on $2d$ environments. Similarly as in the $1d$ case, $\Psi(r)$ is composed of a term diverging with $\log{N}$ (not depending on $r$), and a functional part in $r$ (not depending on $N$). In deriving Eq.~\eqref{coll_dens_free_2d} we have neglected a constant term which is small compared to $\log(N)$, namely, $[\log(2)-\gamma/2]/\pi$, where $\gamma \simeq 0.57721$ is the Euler's gamma constant~\cite{abramowitz}.

The collision density with leakages at $r=r_e$ can be obtained again by the method of images, whence
\begin{equation}
\Psi_R(r) =  \frac{ e^{-r}}{2 \pi r} - \frac{ e^{-r_e}}{2 \pi r_e} + \frac{{\text Ei}\left(- r\right)-{\text Ei}\left(- r_e\right)  -2\log\left(\frac{r}{r_e} \right)}{2\pi },
\label{coll_dens_2d_leak}
\end{equation}
where $r_e=R\left[ 1+u_2/R\right] $ and the Milne's constant $u_2 \simeq 1-2/\pi^2$~\cite{freund}.

The moments associated to the free propagator clearly diverge. For the propagator with leakages at the boundary $r=r_e \ge R$ the mean first-passage time within a sphere of radius $R$ reads
\begin{eqnarray}
\langle t_R \rangle = \frac{1 - e^{-R} + Re^{-R} + R^2- \frac{R^2 e^{-r_e}}{r_e}}{2}\tau +\nonumber \\
 + \frac{R^2 {\text Ei}\left(- R\right) -  R^2 {\text Ei}\left(- r_e\right) - 2 R^2 \log(\frac{R}{r_e})}{2}\tau.
\end{eqnarray}
When $R \gg1$, we have $r_e \simeq R$, and we get $\langle t_R \rangle \simeq (1+R^2)\tau/2$ in the diffusion limit.

As for the collision density with absorptions, calculations analogous to the $1d$ case lead to
\begin{equation}
\Psi_a(r)=\frac{pK_0\left(r\sqrt{1-p^2} \right) }{\pi}+\frac{1}{2\pi}\int_{0}^{+\infty}\frac{zJ_0\left(rz \right)}{\sqrt{1+z^2}+p}dz. 
\end{equation}
We could not find an explicit expression for the latter integral in terms of elementary functions. However, the limits for small and large scattering probability can be easily obtained, and read
\begin{equation}
\Psi_a(r) \simeq \frac{e^{-r}}{2\pi r}+\frac{pK_0\left(r \right) }{\pi},
\end{equation}
when $p \to 0$, and
\begin{equation}
\Psi_a(r) \simeq \frac{\log\left( \frac{1}{1-p}\right) }{2 \pi} +\frac{e^{-r}}{2\pi r} +\frac{{\text Ei}\left(- r\right)-2\log(r)}{2\pi},
\end{equation}
when $p \to 1$, respectively. The former expression gives the uncollided propagator at the leading order, whereas the latter diverges logarithmically as $p \to 1$.

\subsection{Three-dimensional setup, $d=3$}
\label{threedimension}

The case $d=3$ plays a prominent role, among others, in Reactor Physics, in that it concerns the transport of neutrons and photons through matter~\cite{bell, wigner, cercignani}, and is key in describing electron transport in bulk semiconductors~\cite{jacoboni_book}. On the basis of the strikingly similar form of Eqs.~\eqref{1d_propagator} and~\eqref{2d_propagator}, it would be tempting to postulate an analogous expression for the propagator in $3d$. For $d=1$ we have indeed the functional form $\Psi(r|n) \propto r^{-1/2+n}K_{-1/2+n}(r)$, and for $d=2$ $\Psi(r|n) \propto r^{-1+n/2}K_{-1+n/2}(r)$. Then we could conjecture an exponent $-d/2+n/d$, so that
\begin{equation}
\Psi^*(r|n) = \frac{r^{-1/2+n}K_{-1/2+n}(r)}{2^{\frac{1}{2}+\frac{n}{3}}\pi^{\frac{3}{2}}\Gamma\left( \frac{n}{3}\right) },
\label{approximate_3d} 
\end{equation}
by imposing normalization. Unfortunately, this is not the case, and $\Psi^*(r|n)$ is not the true $3d$ propagator. Actually, few explicit results can be derived, and much of the analysis is therefore devoted to the asymptotic behavior. The transition kernel reads
\begin{equation}
\pi_3(\ell)=\frac{ e^{-\ell}}{4\pi \ell^2 },
\end{equation}
whose Fourier transform is
\begin{equation}
\pi_3(z)=\frac{\arctan\left(z  \right) }{z }.
\end{equation}
The propagator $\Psi(r|n)$ with initial condition $\Psi(z,0)=1$ involves then the following integral
\begin{equation}
\Psi(r|n)=\frac{1}{2\pi^2 r}\int_{0}^{+\infty}z \sin\left( r z\right) \left[  \frac{\arctan\left(z  \right) }{z }\right]^n dz,
\end{equation}
which can not be carried out explicitly for arbitrary $n$. In the diffusion limit $z\ll 1$ we have $[\arctan(z)/z]^n \simeq 1- nz^2/3$, so that
\begin{equation}
\Psi(r|n) \simeq \frac{3\sqrt{3}e^{-\frac{3r^2}{4n}}}{8n^{3/2}\pi^{3/2}},
\label{diff_3d}
\end{equation} 
as expected from Eq.~\eqref{diff_eq_n}. In Fig.~\ref{figpn_d3} we compare the Monte Carlo simulation results (symbols) with the diffusion limit, Eq.~\eqref{diff_3d} (dashed lines), and with the approximate propagator, Eq.~\eqref{approximate_3d} (solid lines). Remark that Eq.~\eqref{approximate_3d} provides a fairly accurate approximation of the simulation results, except close to the source.

After carrying out the sum over $n$, the collision density $\Psi(r)$ is given by the following integral 
\begin{equation}
\Psi(r)=\frac{1}{2\pi^2 r}\int_{0}^{+\infty}z \sin\left( r z\right) \frac{\arctan\left(z  \right) }{z -\arctan\left(z  \right)}dz,
\end{equation}
which again can not be performed explicitly~\cite{case}. As before, we consider then the asymptotic behavior. Denoting $h(z)= \arctan(z) /[z-\arctan(z)]$, we have
\begin{equation}
h(z) \simeq \frac{3}{z^2}+\frac{4}{5}-\frac{36}{175}z^2+...
\end{equation}
in the diffusion limit $z \ll 1$, and
\begin{equation}
h(z) \simeq \frac{\pi}{2z}+\left(\frac{\pi^2-4}{4z^2} \right)+ \left(\frac{\pi^3-8\pi}{8z^3} \right)+...
\end{equation}
close to the source. Similar expansions appear, e.g., in~\cite{case}, as derived from the analysis of the Boltzmann equation. Correspondingly, by performing the respective integrations we have
\begin{equation}
\Psi(r) \simeq \frac{3}{4 \pi r }
\label{far_source_3d}
\end{equation}
for $r \gg 1$, i.e., far from the source, and
\begin{equation}
\Psi(r) \simeq \frac{1}{4\pi r^2 } +\left(\frac{\pi^2-4}{16\pi r } \right)+\left(\frac{8-\pi^2}{16\pi}\right)\left[ \log\left( r\right)+\gamma-1 \right] + ...
\label{close_source_3d}
\end{equation}
for $r \ll 1$, i.e., close to the source. Remark that for $d=3$ $\Psi(r)$ does not diverge even for infinite domains without absorptions. In Fig.~\ref{figpsi_3d} we compare the Monte Carlo simulation results (symbols) with the asymptotic limits close to and far from the source, Eqs.~\eqref{close_source_3d} and~\eqref{far_source_3d}, respectively (dashed lines). The simulation results progressively approach the asymptotic limits as the number $N$ of summed collisions increases.

Eqs.~\eqref{close_source_3d} and~\eqref{far_source_3d} can provide asymptotic estimates for the collision density with leakages at the boundary $r=R$. By the method of images, we have that the collision density with boundaries is $\Psi_R(r)=\Psi(r)-\Psi(r_e)$, with $r_e=R\left[ 1+u_3/R\right] $ and $u_3 \simeq 0.7104$~\cite{placzek}.

The moments of the residence time within a sphere of radius $R$ can be explicitly computed based on Eq.~\eqref{average_n_t} for the free propagator $\Psi(r|n)$, i.e., when particles can freely cross the surface of the sphere. We have
\begin{equation}
\langle t_R \rangle \simeq \frac{3}{2}R^2\tau
\end{equation}
when $R \gg 1$, and
\begin{equation}
\langle t_R \rangle \simeq \left( R+ \frac{\pi^2-4}{8}R^2\right) \tau
\end{equation}
for $R \ll 1$. Moreover, for leakage boundary conditions at the surface, from $\Psi_R(r|n)$ we have
\begin{equation}
\langle t_R \rangle \simeq \frac{R^2}{r_e}\left(\frac{3}{2}r_e -R\right)\tau
\end{equation}
when $R \gg 1$, and
\begin{equation}
\langle t_R \rangle \simeq \left( R+R^2\frac{\pi^2-4}{8}+\frac{R^2}{r_e^2}\frac{r_e(4-\pi^2)-4}{12}\right) \tau
\end{equation}
for $R \ll 1$.

\subsection{Four-dimensional setup, $d=4$}
\label{fourdimension}

The case $d=4$ is briefly presented here for the sake of completeness. The transition kernel reads
\begin{equation}
\pi_4(\ell)=\frac{ e^{-\ell}}{2\pi^2 \ell^3},
\end{equation}
whose Fourier transform is
\begin{equation}
\pi_4(z)=\frac{2}{1+\sqrt{1+z^2 }}.
\end{equation}
We could not find an explicit representation for the inverse Fourier transform of $\Psi(z|n)=\pi_4(z)^n$. Nonetheless, the propagator $\Psi(r,t|n)$ is known and reads
\begin{equation}
\Psi(r,t|n)=\frac{n}{\Gamma\left(n-1 \right) }\frac{e^{-vt}}{\pi^2 (vt)^{1+n}}\left[(vt)^2-r^2\right]^{n-2}
\end{equation}
for $vt \ge r$~\cite{paasschens}. Hence, it follows that the propagator $\Psi(r|n)=\int\Psi(r,t|n)dt/\tau$ can be obtained from solving the integral
\begin{equation}
\Psi(r|n)=\frac{n}{\pi^2\Gamma\left(n-1 \right)}\int^{+\infty}_{r}e^{-z}z^{-1-n}\left(z^2-r^2 \right)^{n-2}dz.
\end{equation}
This integral can be performed, and gives
\begin{equation}
\Psi(r|n)=\frac{1}{2^4\pi^{3/2}}\left[A + B +C \right] ,
\label{propagator_4d}
\end{equation}
where
\begin{eqnarray}
&A=-n2^n\frac{ \Gamma(2-n)}{\Gamma\left(\frac{5-n}{2}\right)\Gamma\left(\frac{6-n}{2}\right)}~_1F_2\left( 2-n,\frac{5-n}{2},\frac{6-n}{2};\frac{r^2}{4}\right) ,\nonumber\\
&B=-2^4r^{n-4}\frac{\Gamma\left(\frac{2-n}{2} \right) }{\Gamma\left( \frac{1}{2}\right) \Gamma\left( \frac{-2+n}{2}\right)}~_1F_2\left(-\frac{n}{2},\frac{1}{2},\frac{-2+n}{2};\frac{r^2}{4}\right),\nonumber\\
&C=2^2nr^{n-3}\frac{\Gamma\left(\frac{1-n}{2} \right) }{\Gamma\left( \frac{3}{2}\right)\Gamma\left( \frac{n-1}{2}\right) }~_1F_2\left(\frac{1-n}{2},\frac{3}{2},\frac{-1+n}{2};\frac{r^2}{4}\right),
\end{eqnarray}
$_1F_2()$ being an hypergeometric function~\cite{abramowitz}.

As for the collision density $\Psi(r)$, we have
\begin{equation}
\Psi(r)=\frac{1}{2\pi^{2}r}\int_0^{+\infty} J_{1}(rz) \left(1+\sqrt{1+z^2} \right) dz,
\end{equation}
which can be computed explicitly and gives
\begin{equation}
\Psi(r)=\frac{ e^{-r}}{2 \pi^2 r^3}+\frac{1}{\pi^2 r^2}.
\label{coll_density_4d_free}
\end{equation}

Finally, the collision density in presence of leakages at the boundary $r=R$ can be obtained by resorting to the method of images, whence $\Psi_R(r)=\Psi(r)-\Psi(r_e)$, with $r_e=R\left[ 1+u_4/R\right] $. The constant $u_4$ has been estimated by running a Monte Carlo simulation and determining the extrapolation length, and reads $u_4 \simeq 0.5$.

The moments of the residence time within a sphere of radius $R$ can be explicitly computed based on Eq.~\eqref{average_n_t} for the free propagator $\Psi(r|n)$, namely,
\begin{equation}
\langle t_R \rangle = \left( 1+R^2-e^{-R}\right)\tau .
\end{equation}
Moreover, for leakage boundary conditions at the surface, from $\Psi_R(r|n)$ we have
\begin{equation}
\langle t_R \rangle = \left(1+R^2-e^{-R}-\frac{R^4 e^{-r_e}}{4r_e^3}-\frac{R^4}{2r_e^2}\right)\tau.
\end{equation}

\section{Conclusions}
\label{conclusions}

In this paper, we have examined the dynamics of exponential flights and their relation with the linear Boltzmann equation, a subject that arises in many areas of Physics or Biology. In particular, we have focused on $i)$ the propagator $\Psi(r|n)$, which describes the ensemble evolution of the transported particles as a function of the number of collisions, and $ii)$ the collision density $\Psi(r)$, which is related to the particle equilibrium distribution. Moreover, the connection between the number of collisions and time has been examined. We have provided the framework for a generic $d$-dimensional setup, which allows emphasizing the key role of $d$ in determining the properties of $\Psi(r|n)$ and $\Psi(r)$.

In this context, we have shown that knowledge of $\Psi(r)$ formally allows deriving the moments of the residence time (or equivalently of the collision number) within a given volume, which is key in assessing many physical properties of the system at study. The role of boundary conditions has been explored by considering leakages from a given domain, via the method of images. The behavior of the spatial moments of the particle ensemble has been examined, as well.

We have then provided specific results for one-speed isotropic transport in infinite as well as bounded domains, and for absorbing or purely scattering media. The case $d=1$ has been considered as a prototype model of exponential flights along a straight line, where only two directions (forward or backward) are possible. Due to this simplification, most quantities can be explicitly derived. The case $d=2$ has been analyzed in detail: despite the calculations being non trivial, in some cases closed-form results can be obtained. In particular, we have provided an expression for the collision density, which, coupled with the method of images, might be useful for a realistic description of migration on bounded surfaces. The case $d=3$ is key in most real-world applications, such as the propagation of neutrons or photons in matter, or electrons in bulk semiconductors. Unfortunately, this case turns out to be hardly amenable to closed-form analytical formulas, and most results concern the asymptotic behavior of the particles, either close to or far from the source. Finally, the case $d=4$ has been considered for the sake of completeness. Moreover, Monte Carlo simulations have been performed so as to validate the proposed results and support the analysis of the asymptotic behavior. A good agreement is found between theoretical predictions and numerical simulations.

By virtue of the increasing power of Monte Carlo methods in solving realistic three-dimensional transport problems, one might argue that finding closed-form results for simple systems has a limited interest. However, we are persuaded that analytical and asymptotic formulas may turn out to be useful, in that they can help in improving Monte Carlo algorithms by a clever use of the relation existing between different variables. Moreover, as exponential flights are a transversal field, the cross-over between distinct areas of science might hopefully shed some light at achieving long-standing goals in transport theory, such as full analytical solutions for $3d$ systems.

\acknowledgments The authors wish to thank Drs.~F.~Malvagi and A.~Rosso for careful reading of the manuscript and useful discussions.

\appendix
\section{Spatial moments}
\label{appendix a}

We begin by computing the $m$-th coefficient of the Taylor expansion of $z^{1-d/2} J_{d/2-1}(r z)$ with respect to $z$, which reads
\begin{equation}
\frac{1}{m!} \frac{\partial^m}{\partial z^m}\left[ z^{1-d/2} J_{d/2-1}(r z)\right]_{z=0}=\frac{i^m \left( \frac{r}{2}\right)^{\frac{d}{2}+m-1} }{\Gamma\left( 1+\frac{m}{2}\right)  \Gamma\left(\frac{d+m}{2} \right) },
\end{equation}
for even $m$, and zero otherwise. Apply now the $m$-th derivative to a function $f(z)$ such that $f(r)$ has a spherical symmetry. Recalling then the definition of the moment $\langle r^m \rangle$ from Eq.~\eqref{mom_int}, we have
\begin{equation}
\frac{1}{m!} \frac{\partial^m}{\partial z^m}\left[ f(z)\right]_{z=0}=\frac{i^m 2^{1-\frac{d}{2}-m} (2\pi)^{\frac{d}{2}}}{\Gamma\left( 1+\frac{m}{2}\right)  \Gamma\left(\frac{d+m}{2} \right) }\frac{\langle r^m \rangle}{\Omega_d}.
\end{equation}
Rearranging the coefficients, we can finally express the spatial moments of $f(r)$ in terms of the $m$-th $z$-derivative of $f(z)$, namely,
\begin{equation}
\langle r^m \rangle=\frac{\sqrt{\pi}\Gamma\left( \frac{d+m}{2}\right) }{\Gamma\left( \frac{d}{2}\right)\Gamma\left( \frac{1+m}{2}\right)} \frac{\partial^m}{\partial (iz)^m}  \left[ f(z)\right]_{z=0}.
\end{equation}

\section{Collision density for $d=2$}
\label{appendix b}

In order to find the collision density associated to the free propagator, we begin by decomposing the sum over the collisions $n$ into even and odd index, i.e.,
\begin{equation}
\Psi(r)=\lim_{N \to \infty}\left[ \sum_{n\, {\text even}}\Psi(r|n)+ \sum_{n\, {\text odd}}\Psi(r|n) \right].
\end{equation}
Then, by remarking that for even $n$
\begin{equation}
\Psi_{\text even}(z)=\sum_{n\, {\text even}}\Psi(z|n) = \frac{ 1 - \left( 1 + z^2\right)^{-N} }{z^2},
\end{equation}
we get
\begin{equation}
\Psi_{\text even}(r) \simeq \frac{ \log(\sqrt{N}) - \log(r) }{2 \pi}
\end{equation}
for large $N$. We have neglected a constant term of the kind $[\log(2)-\gamma/2]/(2\pi)$, which is small compared to $\log(N)$.

For odd $n$, we can use the series representation
\begin{eqnarray}
&\Psi(r|n_{\text odd})= \sum_{k\, {\text even}} \frac{2^{-2-k}\Gamma\left(2-\frac{n}{2} \right) \Gamma\left(-1+\frac{n}{2} \right)}{\pi \Gamma\left( 1+\frac{k}{2}\right) \Gamma\left( 2+\frac{k}{2}-\frac{n}{2}\right) \Gamma\left( \frac{n}{2}\right) }r^k +  \nonumber \\
&\sum_{\substack{(n-k) \, {\text even}\\k\ge n-2}}\frac{2^{-2-k}\Gamma\left(2-\frac{n}{2} \right)}{\pi \Gamma\left( 1+\frac{k}{2}\right) \Gamma\left( 2+\frac{k}{2}-\frac{n}{2}\right) }r^k.
\label{double_sum}
\end{eqnarray}
Now, carrying out the double sum over odd $n$ and over $k$, from Eq.~\eqref{double_sum} we get
\begin{equation}
\Psi_{\text odd}(r) \simeq \frac{\log(\sqrt{N}) + \frac{e^{-r}}{r} + {\text Ei}\left(- r\right) -\log(r)}{2\pi},
\end{equation}
for large $N$. Again, to obtain this result we have neglected a constant term of the kind $[\log(2)-\gamma/2]/(2\pi)$, which is small compared to $\log(N)$.

Hence, by summing up we finally obtain
\begin{equation}
\Psi(r) \simeq \frac{\log\left( N\right)+ \frac{e^{-r}}{r} + {\text Ei}\left(- r\right) -2\log\left(r \right)}{2\pi}.
\end{equation}

\end{document}